%% file: main.tex
\documentclass[conference]{IEEEtran}
\usepackage{fancyhdr}
\usepackage{cite}
\usepackage{amsmath,amssymb,amsfonts}
\usepackage{algorithmic}
\usepackage{graphicx}
\usepackage{textcomp}
\usepackage[dvipsnames,table,xcdraw]{xcolor}
\usepackage{subcaption}
\usepackage{multirow}
\usepackage{url}
\usepackage{mathtools}
\usepackage{wasysym}
\usepackage{nicefrac}
\usepackage{hyperref}
\usepackage{array}

\def\BibTeX{{\rm B\kern-.05em{\sc i\kern-.025em b}\kern-.08em
    T\kern-.1667em\lower.7ex\hbox{E}\kern-.125emX}}
\begin{document}

\title{LLM-Pilot: Characterize and Optimize Performance\\of your LLM Inference Services}

\author{\IEEEauthorblockN{Malgorzata Lazuka}
\IEEEauthorblockA{\textit{IBM Research, ETH Zurich}\\
Zurich, Switzerland \\
mal@zurich.ibm.com}
\and
\IEEEauthorblockN{Andreea Anghel}
\IEEEauthorblockA{\textit{IBM Research}\\
Zurich, Switzerland \\
aan@zurich.ibm.com}
\and
\IEEEauthorblockN{Thomas Parnell}
\IEEEauthorblockA{\textit{IBM Research}\\
Zurich, Switzerland \\
tpa@zurich.ibm.com}
}

\newcommand{\greencheck}{{\color{Green}\checkmark}}
\newcommand{\redcross}{{\color{Red}$\times$}}
\newcommand{\yellowdash}{{\color{blue}--}}
\DeclarePairedDelimiter{\ceil}{\lceil}{\rceil}

\maketitle
\thispagestyle{fancy}
\lhead{}
\rhead{}
\chead{}
\lfoot{}
\rfoot{}
\cfoot{\footnotesize{\copyright 2024 IEEE. Personal use of this material is permitted. Permission from IEEE must be obtained for all other uses, in any current or future media, including reprinting/republishing this material for advertising or promotional purposes, creating new collective works, for resale or redistribution to servers or lists, or reuse of any copyrighted component of this work in other works.}}
\renewcommand{\headrulewidth}{0pt}
\renewcommand{\footrulewidth}{0pt}

\input{0-abstract}


\input{1-introduction}

\input{2-background.tex}

\input{3-characterization.tex}

\input{4-prediction.tex}

\input{5-evaluation.tex}

\input{6-related_works.tex}

\input{7-conclusion.tex}

\input{acknowledgement}

\bibliographystyle{plain}
\bibliography{citations}

\end{document}

%% file: 0-abstract.tex
% !TEX root = main.tex
\begin{abstract}
As Large Language Models (LLMs) are rapidly growing in popularity, LLM inference services must be able to serve requests from thousands of users while satisfying performance requirements. 
The performance of an LLM inference service is largely determined by the hardware onto which it is deployed, but understanding of which hardware will deliver on performance requirements remains challenging.
In this work we present LLM-Pilot -- a first-of-its-kind system for characterizing and predicting performance of LLM inference services.
LLM-Pilot performs benchmarking of LLM inference services, under a realistic workload, across a variety of GPUs, and optimizes the service configuration for each considered GPU to maximize performance. 
Finally, using this characterization data, LLM-Pilot learns a predictive model, which can be used to recommend the most cost-effective hardware for a previously unseen LLM.
Compared to existing methods, LLM-Pilot can deliver on performance requirements 33\% more frequently, whilst reducing costs by 60\% on average.
\end{abstract}

%% file: 1-introduction.tex
% !TEX root = main.tex
\section{Introduction}
\label{sec:intro}

Large Language Models (LLMs) have gained massive popularity in recent years, in both industry and the research community \cite{zhao2023survey} for their remarkable capability of performing a wide variety of natural language processing tasks \cite{naveed2024comprehensive}.
In particular, LLMs excel at generating text and programming code, conducting conversation as chatbots, extracting information from text, and language translation.
The development of LLMs is only accelerating and considered to be a modern-day Moore's law \cite{MooresLaw, bommasani2022opportunities}, with numbers of parameters of newly released LLMs growing exponentially \cite{AssemblyAIBlog, narayanan2021efficient, wei2022emergent}.
This is largely due to the fact that, until now, training larger LLMs has been a sure path to improving the LLM's output quality \cite{NEURIPS2020_1457c0d6, kaplan2020scaling}.
Unfortunately, growing LLMs in size leads to the necessity to scale the hardware used for their training and inference~\cite{bommasani2022opportunities, strubell2019energy}.
As most research efforts in this area go into improving the performance of LLM training, optimizing the performance of LLM inference services remains a fairly unexplored area \cite{samsi2023words}.
Therefore, we are starting to observe a new challenge emerging: \emph{Once you have trained a state-of-the-art LLM with billions of parameters, how do you deploy it in a cost-effective way while ensuring sufficient performance?}

The choice of hardware to which the inference service is deployed can significantly impact the resulting performance~\cite{samsi2023words}.
As LLM inference services must store billions of model weights, the choice of hardware is limited to a range of powerful GPUs, which are currently scarce and in all-time high demand \cite{shortage}.
As a result, GPUs are both hard to get and hard to afford.
While cloud providers offer on-demand GPU instances, in the long-term they are more expensive than buying GPUs and institutions often prefer to pool on-prem resources instead \cite{strubell2019energy}.
In this work we consider the perspective of such an institution, which owns and maintains a large cluster of heterogeneous GPUs. 
Each GPU has some cost per unit time associated with it, which may be related to the total cost of ownership, energy consumption or some combination thereof. 
We consider two roles with respect to this cluster: a cluster user and a cluster administrator. 
The user wishes to deploy an LLM inference service and has a set of performance requirements that should be met, while the administrator has an interest that each inference service incurs as little cost as possible, thus ensuring that the resources are efficiently utilized. 
In some private cloud-like scenarios, the burden of cost may also be passed onto the user, who may be billed internally for the resources consumed. 
Trying to satisfy both performance and cost requirements is challenging since the performance of a given LLM inference service on a given GPU in the cluster is a-priori unknown. 
Additionally, it is impractical for the user to run a large set of experiments to characterize performance across different GPUs, because the cluster is typically close to fully-utilized. 
This motivates the two problems considered in this work: (1) how the administrator can collect data offline to characterize the performance of different LLMs on different GPUs and (2) how this data can assist the user to make online decisions about the type and number of GPUs that will satisfy their performance requirements in the most cost-effective way. 

In this work we present LLM-Pilot, a first-of-its-kind system for characterizing the performance of LLM inference services, and recommending the cheapest deployment for a new LLM such that its performance requirements are met.
LLM-Pilot consists of two main components: the performance characterization tool and the GPU recommendation tool.
The performance characterization tool can be used offline, by the cluster administrator, to benchmark the performance of a collection of LLM inference services across the GPUs of the cluster.
While doing so, it ensures that the inference service is subject to a realistic load of inference requests and that the server-side batching algorithm has been configured individually for each GPU. 
The output is a characterization dataset containing performance measurements for a variety of LLMs across the different GPUs of the cluster. 
The GPU recommendation tool provides the cluster user with online recommendations regarding how to deploy an unseen LLM in the most cost-effective way whilst satisfying their performance requirements.
This is achieved by learning a predictive performance model, fitted to the offline characterization data, and tailored to the user's specific performance requirements. 
In summary, the contributions of this work are as follows:
\begin{enumerate}
	\item{An LLM inference benchmarking tool\footnote{\label{repo1}Available at: \url{https://github.com/fmperf-project/fmperf}.} which ensures that the services are optimized for the specific hardware, and internally uses our novel workload generator\textsuperscript{\ref{repo1}}, which subjects the services to a realistic workload of inference requests based on a large collection of production traces.}
	\item{A performance dataset comparing the inference performance of many LLMs running on a variety of GPUs, which we have open sourced in order to deepen the understanding of LLM inference performance and its dependence on the GPU choice\footnote{\label{repo2}Available at: \url{https://github.com/IBM/LLM-performance-prediction}.}.}
	\item{A GPU recommendation tool\textsuperscript{\ref{repo2}} which can ensure satisfying the performance requirements 33\% more frequently than state-of-the-art methods, and reduces cost of recommended type and number of GPUs by 60\% on average, thanks to the use of a novel performance prediction model designed specifically for this use-case.}
\end{enumerate}
The structure of this work is as follows: in Sec. \ref{sec:background} we discuss the background on the performance and deployment of LLM inference services.
Then, we present the performance characterization tool (Sec. \ref{sec:characterization}) and the GPU recommendation tool (Sec. \ref{sec:prediction}) of LLM-Pilot.
In Sec. \ref{sec:evaluation} we evaluate various novel components of LLM-Pilot, and in Sec. \ref{sec:related_works} we discuss prior works on related topics.

%% file: 2-background.tex
% !TEX root = main.tex
\section{Background}
\label{sec:background}

\subsection{LLM inference performance requirements}
Unlike inference services based on classical machine learning or deep learning models, LLMs process requests in two phases.
First, in the so-called \emph{prompt processing} phase, they split the input text into smaller segments called tokens, process them and populate the key-value (KV) cache \cite{vLLM}, and finally generate the first output token.
Then, in the \emph{decode} phase, they update the KV cache and sequentially generate the output tokens, which are later converted into an output text and sent to the user.
The latency of these phases (and consequently the end-to-end latency) can vary significantly depending on the number of input and output tokens.
Therefore, in case of LLMs one typically defines two latency metrics.
Time to first token (TTFT) measures the overhead of generating the first output token resulting from processing the input tokens.
The inter-token latency (ITL) measures the latency of generating each subsequent output token.
Depending on the application, one of these metrics can have larger impact on the end user's experience than the other.
For example, in case of LLM-powered chatbots, it is important for the user's experience that the output starts being generated quickly, while the speed of subsequent tokens does not need to exceed the human reading speed \cite{goodput}.
On the other hand, when LLMs are used for text summarization, longer initial overhead is acceptable as long as long output texts can be generated quickly \cite{goodput}.
Inference services are typically subject to a service-level agreement (SLA) which outlines the requirements regarding the service's performance \cite{SLA}.
In case of LLM inference services, however, one must ensure that both TTFT and ITL meet their respective SLA constraints.

\subsection{LLM inference servers}
In practice, in order to perform inference on a trained LLM, one uses a framework called an inference server (e.g., vLLM~\cite{vLLM}, TGIS~\cite{TGIS}, TGI~\cite{TGI}, Orca~\cite{Orca1}, NVIDIA Triton Inference Server~\cite{triton}), which acts as a bridge between the LLM and the users sending their inference requests. 
The inference server handles running all incoming requests through the LLM and sending back the generated responses.
As the incoming inference requests can strongly vary in terms of number of input and output tokens, inference servers typically use \emph{continuous batching} \cite{Orca1} to improve utilization of the hardware hosting the inference service.
In continuous batching, the server maintains a single batch of requests from various users that are being processed at any given moment.
When processing of certain requests in the batch finishes, new requests are introduced into the batch from the queue, while other requests in the batch continue processing.
This way, requests with diverse numbers of input and output tokens can be processesed in parallel but small requests can be processed quickly, without waiting for larger requests in the batch to finish processing.

When an LLM is deployed on a GPU, a large amount of its capacity is used for storing the LLM itself.
The remaining capacity is required for storing the KV cache associated with the batch of requests currently being processed, as well as any auxiliary data structures required to perform a forward pass with the LLM.
The larger the storage space assigned to the batch, the more requests can be processed in parallel, which in turn improves the throughput of the inference service.
Inference servers control the maximum size of the batch using equivalent parameters under different names, e.g., max batch weight in TGIS \cite{TGIS}, max num batch tokens in vLLM \cite{vLLM}, and max batch total tokens in TGI \cite{TGI}.
In this work, we refer to it as the \emph{maximum batch weight}.
The maximum batch weight is the maximum allowed `volume' of the batch, defined as the total number input and output tokens of all requests processed in the batch at any given time.
This indirectly determines the maximum storage space that can be occupied by the batch and significantly impacts the inference performance.

\begin{figure}[t]
\centerline{\includegraphics[width=\columnwidth]{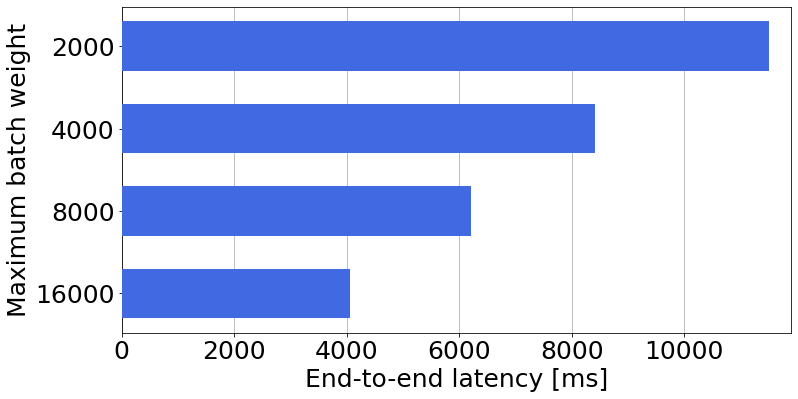}}
\caption{Median end-to-end latency achieved by the inference service of a selected LLM (bigcode/starcoder \cite{starcoder}) deployed on one A100 GPU with varying maximum batch weight, for 128 concurrent users.}
\label{fig:batch_weight_bar}
\end{figure}

\begin{figure}[t]
\centerline{\includegraphics[width=\columnwidth]{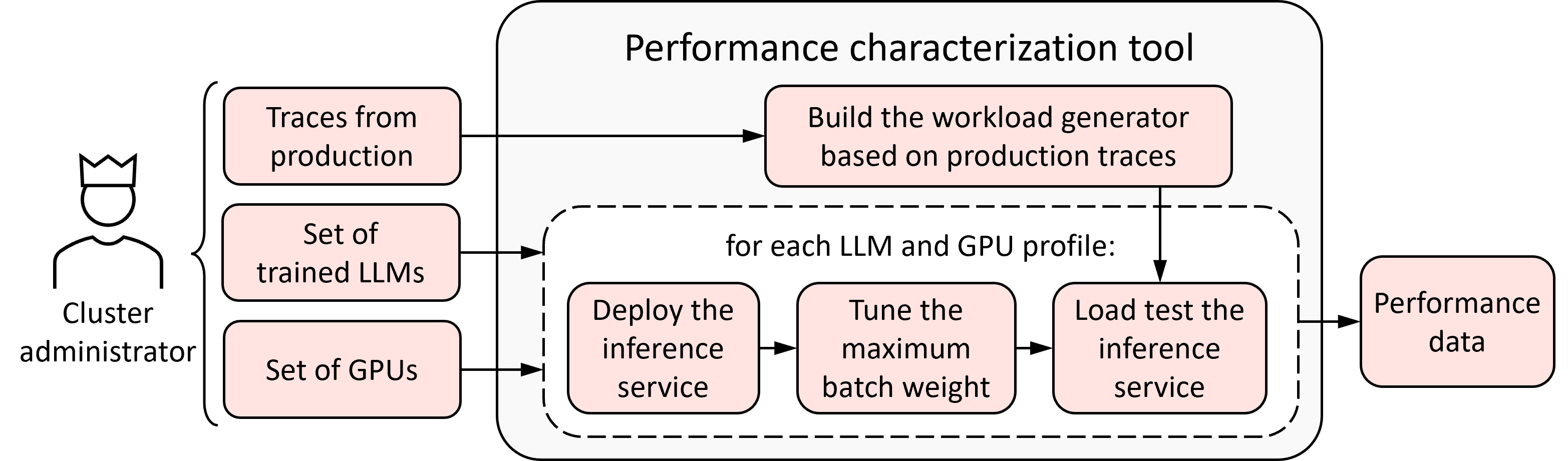}}
\caption{Architecture of the performance characterization tool.}
\label{fig:benchmarking_structure}
\end{figure}

Tuning the maximum batch weight is critical to achieving optimal inference performance.
For example, when the maximum batch weight is doubled, we can expect the queuing time of requests to decrease because twice as many tokens are being processed in parallel.
However, the end-to-end latency comprises both the queuing time and the processing time (i.e., time needed to generate all output tokens).
If doubling the maximum batch weight increases the processing time of each request by more than 2$\times$, the final end-to-end latency will actually be worse.
Therefore, in Fig.~\ref{fig:batch_weight_bar} we analyze the end-to-end latency of a selected LLM inference service running on the same GPU with varying maximum batch weight.
The chart shows that increasing the maximum batch weight improves the end-to-end latency: the largest maximum batch weight results in approx. 2.8$\times$ lower end-to-end latency than the smallest one.
This confirms that (a) the maximum batch weight strongly impacts the inference performance, and (b) in order to optimize performance under the load of inference requests from our workload generator (see Sec.~\ref{subsec:workload_generator}), the maximum batch weight should be set as high as possible.

\subsection{Deploying LLM inference services}
When an inference service is to be used in production, it is deployed onto a Kubernetes (k8s) \cite{k8s} or Openshift \cite{Openshift} cluster via an abstraction called a \emph{Deployment}.
Each deployment manages a number of replicas of the inference service, and each replica is deployed to the cluster via an abstraction called a \emph{Pod}.
Load balancing is performed across the pods of the deployment, which operate independently, and the number of pods can be scaled up or down based on demand.

Since inference is embarrassingly parallel at the level of requests, and LLM inference requests can easily take more than 100ms, we expect close-to-perfect scaling of the throughput with respect to the number of pods.
We confirm this experimentally, as depicted in Table \ref{tab:scaling}.
In the experiment, we have tested different numbers of pods of the same LLM inference service running on the same GPU, under different numbers of concurrent users whose requests are distributed across pods.
Results confirm near-perfect scaling -- across cases with the same ratio between the number of concurrent users and number of pods (e.g., 1 pod and 8 users, 2 pods and 16 users, etc.), indicated by the same cell background color, the relative standard deviation of throughputs per pod never exceeds 5\% (2\% on average).

\begin{table}[t]
\setlength{\tabcolsep}{4.5pt}
\caption{Average throughput per pod achieved by a varying number of Llama-2-13b pods, each running on a A100 80GB GPU, for various numbers of concurrent users. Same color of diagonally adjacent cells marks cases with the same ratio between the number of pods and the number of concurrent users.}
\begin{tabular}{|c|c|c|c|c|c|c|c|c|}
\hline
                                                                                 & \multicolumn{8}{c|}{Number of concurrent users}                                                                                                                                                                                                                                                                                                                                                                \\ \cline{2-9} \multirow{-2}{*}{\begin{tabular}[c]{@{}c@{}}Number\\ of pods\end{tabular}} & 1                            & 2                            & 4                             & 8                             & 16                            & 32                            & 64                            & 128                           \\ \hline
1                                                                                & \cellcolor[HTML]{D2E4FF}{47.1} & \cellcolor[HTML]{E1EDFF}{78.1} & \cellcolor[HTML]{E7D6FF}{118.2} & \cellcolor[HTML]{F0E5FF}{174.2} & \cellcolor[HTML]{FFD8D8}{237.8} & \cellcolor[HTML]{FFE6E5}{288.6} & \cellcolor[HTML]{FFE2BF}{314.7} & \cellcolor[HTML]{FFF1E1}{292.6} \\ \hline
2                                                                                & \cellcolor[HTML]{E1FFE1}{23.5} & \cellcolor[HTML]{D2E4FF}{44.4} & \cellcolor[HTML]{E1EDFF}{74.7}  & \cellcolor[HTML]{E7D6FF}{114.3} & \cellcolor[HTML]{F0E5FF}{171.1} & \cellcolor[HTML]{FFD8D8}{229.8} & \cellcolor[HTML]{FFE6E5}{282.7} & \cellcolor[HTML]{FFE2BF}{286.3} \\ \hline
4                                                                                & \cellcolor[HTML]{C7FFC7}{11.6} & \cellcolor[HTML]{E1FFE1}{23.3} & \cellcolor[HTML]{D2E4FF}{43.0}  & \cellcolor[HTML]{E1EDFF}{73.5}  & \cellcolor[HTML]{E7D6FF}{113.0} & \cellcolor[HTML]{F0E5FF}{169.5} & \cellcolor[HTML]{FFD8D8}{231.4} & \cellcolor[HTML]{FFE6E5}{283.4} \\ \hline
8                                                                                & \cellcolor[HTML]{FFFFDB}{5.7}  & \cellcolor[HTML]{C7FFC7}{11.7} & \cellcolor[HTML]{E1FFE1}{22.8}  & \cellcolor[HTML]{D2E4FF}{42.3}  & \cellcolor[HTML]{E1EDFF}{72.6}  & \cellcolor[HTML]{E7D6FF}{112.9} & \cellcolor[HTML]{F0E5FF}{169.2} & \cellcolor[HTML]{FFD8D8}{231.1} \\ \hline
\end{tabular}
\label{tab:scaling}
\end{table}

When defining the deployment specification, one must declare what hardware resources should be allocated to each pod.
These include the number and type of GPUs to be assigned to each pod (which we jointly refer to as the \emph{GPU profile}), as well as the number of CPU cores and amount of memory. 
Note that in the case when the specified GPU profile comprises more than one GPU, the weights of the LLM and all computation will be sharded across the GPUs in a tensor-parallel manner.
Tensor parallel deployments may be preferred when dealing with very large models that are too big to fit in a single GPU, and in some cases may also bring latency benefits \cite{tensor_parallel}.
Note that in this setup, each pod has exclusive access to the GPUs assigned to it, and thus there are no effects related to co-location that need to be considered.

%% file: 3-characterization.tex
% !TEX root = main.tex
\section{Performance Characterization Tool}
\label{sec:characterization}

In this section we present the performance characterization tool, the structure of which is presented in Fig. \ref{fig:benchmarking_structure}.
There are two important aspects of performance characterization that we have considered when developing LLM-Pilot.
Firstly, we ensure that the LLM's performance will be characterized under a \emph{realistic} workload.
Therefore, we have acquired and analyzed a large collection of production traces of real inference requests to a variety of LLMs (Sec. \ref{subsec:traces}).
Then, we have developed our own workload generator (Sec. \ref{subsec:workload_generator}) based on the production traces to ensure that the generated workload resembles the real usage of LLM inference services.
Secondly, the performance characterization tool ensures that the maximum batch weight is \emph{optimized} for each GPU profile.
As shown in Fig. \ref{fig:benchmarking_structure}, on each benchmarked GPU profile LLM-Pilot deploys the inference service, tunes the maximum batch weight, and subjects the service to a series of requests from the workload generator, collecting various performance metrics (Sec. \ref{subsec:load_testing}).
Other considerations taken into account when developing the performance characterization tool of LLM-Pilot have been discussed in Sec. \ref{subsec:others}.

\begin{table}[t]
\caption{Characteristics of the production traces used to develop the workload generator.}
\begin{center}
\begin{tabular}{|l|l|}
\hline
\textbf{Time period} & 5.5 months \\ \hline
\textbf{Number of requests}&17.3M \\ \hline
\textbf{Number of users} & approx. 2500 \\ \hline
\textbf{Number of LLMs} & 24 (with 3B--176B parameters)\\ \hline
\textbf{Range of tokens} & input: 1--4093, output: 1--1500 \\ \hline
\textbf{Batch sizes} & 1--5 \\ \hline
\textbf{Additional parameters} & 33 (e.g., decoding method,  top k, \\
\textbf{describing the requests} & top p, repetition penalty, \\
& length penalty, temperature) \\ \hline
\end{tabular}
\label{tab:traces}
\end{center}
\end{table}

\begin{figure}[t]
  \centering
  \includegraphics[width=\linewidth]{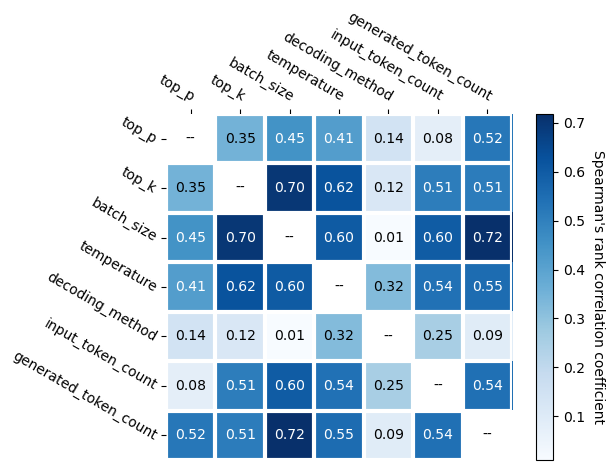}
  \caption{Correlation between selected parameters of requests from the production traces.}
  \label{fig:correlation}
\end{figure}

\subsection{Analysis of production traces}
\label{subsec:traces}

The production traces analyzed in this section come from an LLM inference platform used internally in our organization, which hosts a large number of LLMs deployed on an Openshift cluster running on A100 GPUs, which are made available for many users to send inference requests.
Internally, the platform uses Text Generation Inference Server (TGIS) \cite{TGIS}.
The traces are a record of every inference request (i.e., every request to perform inference on a certain input text using a specific LLM and return the output) sent to the inference platform by every user within a certain period of time.
Each entry in the traces includes the user's id, timestamp and all details of the request (including various TGIS-specific request parameters set by the user), as well as output from the LLM and end-to-end latency of processing the request.
As shown in the detailed characteristics in Table \ref{tab:traces}, the traces were collected over a long period of time and include millions of diverse requests from thousands of users.
To the best of our knowledge, the traces used in this work are larger than any publicly available LLM trace collection published to date \cite{FlanCollection, ShareGPT, ChatbotArena, OpenAssistant}.
Furthermore, the traces were in no way generated or collected synthetically and therefore represent a fully realistic and diverse usage of LLM inference services.

\subsubsection*{Impact on latency}
As shown in Table \ref{tab:traces}, each request in the traces is described by a large number of parameters. 
We analyze the impact of these parameters on the latency through an importance study using a Random Forest (RF) regression model. 
First, we trained a RF regressor on all available trace data to predict the latency of all individual requests using all parameters included in the traces.
The model achieved good accuracy, with the coefficient of determination $R^2 \approx 0.93$.
Then, we evaluated their impact on the RF’s predictions using the Mean Decrease in Impurity (MDI) \cite{breiman1984classification}.
According to our study and our expectation, the parameter with the largest influence on the performance is the number of output tokens, followed by the number of input tokens, the batch size and parameters related to token sampling (decoding method, temperature, top p,  top k).

\subsubsection*{Correlation}
\label{subsubsec:correlation}
Further, we have analyzed the correlation between the request parameters listed above using the Spearman's rank correlation coefficient \cite{Spearman}.
The results presented in Fig. \ref{fig:correlation} indicate that many pairs of parameters are strongly correlated.
Most notably, the parameters with the strongest influence on the performance -- the batch size and the numbers of input and output tokens -- are all strongly correlated with one another.
Therefore, in order for a workload generator to produce realistic workloads, it should preserve the correlation between the parameters characterizing each request.

\subsection{Workload generator}
\label{subsec:workload_generator}

Based on the conclusions drawn in Sec. \ref{subsec:traces}, in each request the workload generator should specify the parameters which impact the latency, and it must consider the fact that in practice some parameters are strongly correlated.
These considerations are an important contribution of our work, as to the best of our knowledge, no prior works on workload generation include any request parameters beyond the batch size and the numbers of input and output tokens \cite{inference-benchmark, vLLM, Optimum, LLMPerf, Fleece, MLPerf}, and in some works the request parameters are treated as independent variables \cite{LLMPerf, Optimum}.

\subsubsection{Modelling the requests}
Internally, the workload generator uses a non-parametric model of requests, which jointly models the distributions of all request parameters.
For each parameter describing the requests, we divide the range of its values into a series of intervals called \emph{bins}.
This allows us to reduce the cardinality of each parameter, as the true parameter values will be replaced with the centers of their respective bin intervals.
For each parameter, we define 64 bins (unless the cardinality of the parameter is lower, in which case we define as many bins as there are unique values).
We aim to define the bins such that they all contain an approximately equal number of requests.
Then, we proceed to create the joint, multi-dimensional model of requests.
Each multi-dimensional bin is defined by a distinct combination of bin assignments for values of all parameters.

\subsubsection{Sampling requests}
Whenever the workload generator needs to produce an inference request, it can draw a random sample from the model of requests by selecting one of its multi-dimensional bins.
The probability of choosing each bin is defined by the histogram of the multi-dimensional bins, i.e. it is proportional to the number of requests from the traces that were assigned to it.
This way, the distribution of drawn samples will be very similar to the empirical distribution observed in the traces.
The final sample is a request with each parameter equal to the center of that parameter's interval in the selected bin.
The input text for the request is generated based on some designated corpus of texts, truncated to match the number of input tokens indicated by the request's parameters.

\subsection{Performance data collection}
\label{subsec:load_testing}

For each LLM and GPU profile to be benchmarked, LLM-Pilot performs a series of actions: (1) it deploys the inference service, (2) tunes the maximum batch weight parameter of the inference server to ensure maximum GPU utilization, and (3) runs a series of load testing experiments using the workload generator, collecting various performance metrics. 
These steps will now be explained in detail.

\subsubsection{Deployment}
\label{subsubsec:deployment}
The tool creates a TGIS deployment on the cluster, using a single pod, and with the number and type of GPUs set according to the given GPU profile. 
The amount of memory available to the pod is set to 250GB, and the number of CPU cores is set to be twice larger than the number of GPUs. 
LLM-Pilot then waits until the pod is created and the LLM has been loaded into GPU memory, before proceeding to the next step. 

\subsubsection{Tuning the batch weight}
\label{subsubsec:inference_service_parameters}

Based on the conclusions drawn in Sec. \ref{sec:background}, we must ensure that the maximum batch weight parameter is set as high as possible to achieve the best possible performance of the inference service.
As GPU profiles vary in terms of memory capacity, so does the highest possible maximum batch weight that can be achieved. 
Thus, in order to be able to compare GPU profiles in a fair way, we must always ensure that the batch weight has been optimized individually for each one.
This is achieved in LLM-Pilot by running a binary search to optimize the maximum batch weight, as an initialization step before starting the inference server.
Namely, in each step of the search, we test a different maximum batch weight value and check if we encounter Out Of Memory (OOM) errors.
This is achieved by passing a sequence of batches to the model that are designed to test all possible corner cases, with respect to the batch size, number of input and output tokens, that can be constructed according to the given maximum batch weight. 
If all corner cases succeed (i.e., none of them result in an OOM error), the batch weight is considered valid, and otherwise it is not.
Once binary search is completed, we take the optimized value for the maximum batch weight and start the inference server.
Once the server is ready to receive requests, we proceed to perform load testing. 

\subsubsection{Load testing}
In each load testing experiment, LLM-Pilot simulates a different number of concurrent users simultaneously sending various requests generated by the workload generator (Sec. \ref{subsec:workload_generator}), for a duration of 2 minutes.
By default, subsequent experiments simulate 1, 2, 4, \ldots, 128 concurrent users, increasing exponentially.
In each experiment, LLM-Pilot logs all generated tokens and the timestamps of their arrival to the client.
From the timestamps, the following performance metrics are extracted:
\begin{itemize}
	\item{time to first token (TTFT) -- the median latency of receiving the first output token. It includes the time spent on queueing and the prompt processing phase.}
	\item{normalized TTFT (nTTFT) -- the median of TTFT latencies of requests divided by their number of input tokens. We create this new metric as its value does not change as significantly as TTFT with the number of input tokens.}
	\item{inter-token latency (ITL) -- the median latency between all subsequent output tokens, excluding the first one.}
	\item{throughput -- the total number of output tokens generated throughout the experiment, divided by its duration.}
\end{itemize}

Once these three steps have been performed for all LLMs and GPUs, all of the collected performance data are aggregated into a characterization dataset, which is discussed in detail in Sec. \ref{subsec:dataset}.
While there exist other tools for benchmarking the performance of LLM inference services (discussed in Sec.~\ref{subsec:benchmarking_tools}), our work is the only one that benchmarks \emph{optimized} inference services by finding the appropriate value of maximum batch weight to maximize the inference performance.

\begin{figure}[t]
  \centering
  \includegraphics[width=\linewidth]{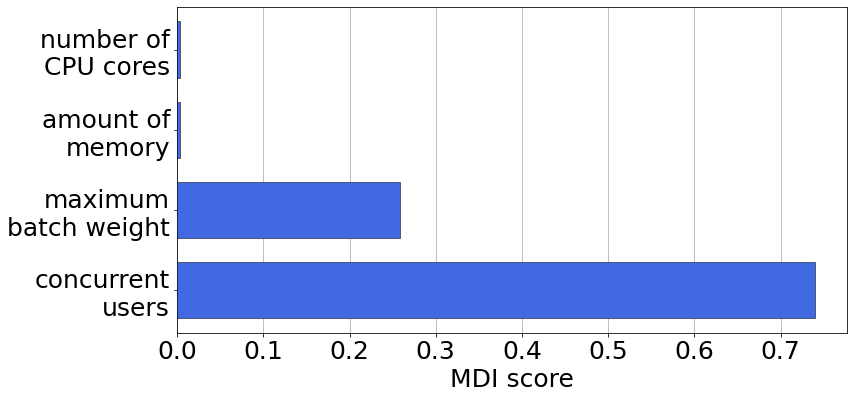}
  \caption{The MDI importance scores of the number of CPU cores, amount of memory, maximum batch weight and number of concurrent users, determined by a RF predicting the TTFT and ITL latency for a selected LLM (bigcode/starcoder \cite{starcoder}).}
  \label{fig:parameter_importances}
\end{figure}

\subsection{Other considerations}
\label{subsec:others}
While we have shown that tuning the maximum batch weight is critical to evaluate performance across different GPUs, a natural question is whether other aspects of the deployment specification also need to be tuned (e.g., the amount of memory and number of CPU cores). 
Similarly to the study in Fig. \ref{fig:batch_weight_bar}, we have run an example LLM's inference service on a single A100 40GB GPU with varying number of CPU cores, amount of memory and maximum batch weight, and measured the inference performance for different numbers of concurrent users as described in Sec. \ref{subsec:load_testing}.
Then, we have trained a RF regressor to predict the TTFT and ITL latencies based on the varying parameters, and performed an importance analysis similarly to Sec. \ref{subsec:workload_generator}.
The MDI importance scores of all parameters are presented in Fig. \ref{fig:parameter_importances}.
The number of CPU cores and memory achieved near-zero scores, over 300$\times$ lower than the MDI score of maximum batch weight.
This suggests that they do not have considerable impact on performance and motivates why LLM-Pilot sets them according to trivial rules.

%% file: 4-prediction.tex
% !TEX root = main.tex
\section{GPU Recommendation Tool}
\label{sec:prediction}

\begin{figure}[t]
\centerline{\includegraphics[width=\columnwidth]{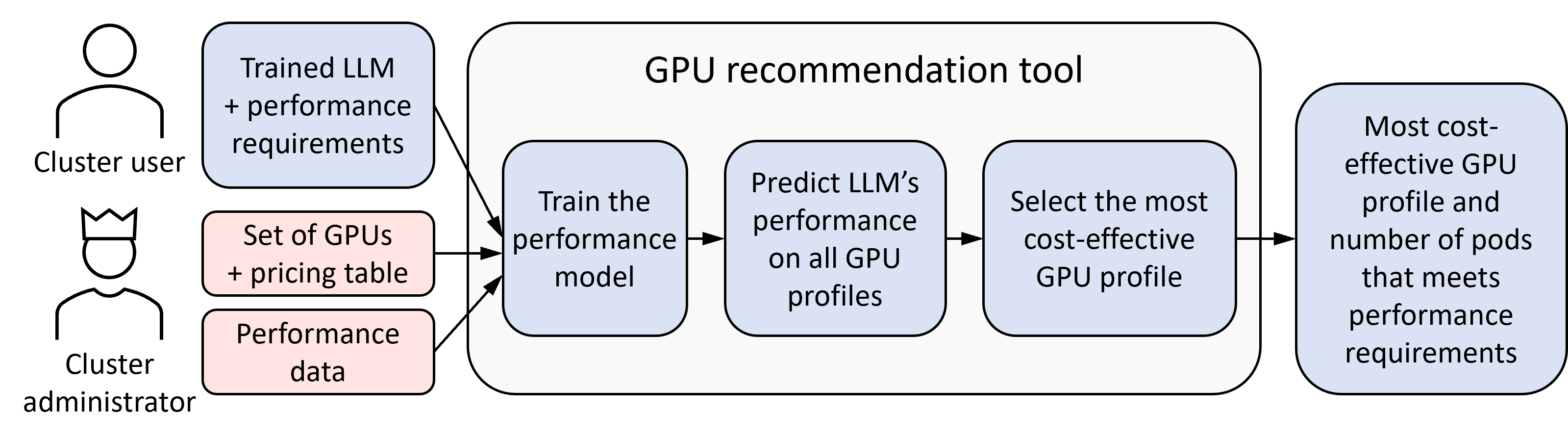}}
\caption{Architecture of the GPU recommendation tool.}
\label{fig:recommendation_structure}
\end{figure}

In this section, we describe LLM-Pilot's GPU recommendation tool, depicted schematically in Fig. \ref{fig:recommendation_structure}.
In Sec \ref{subsec:problem_statement} we define the problem that the GPU recommendation tool aims to solve, and in Sec. \ref{subsec:solution} we describe how LLM-Pilot solves it.
LLM-Pilot's ability to recommend GPUs for unseen LLMs has been evaluated in Sec. \ref{subsec:evaluation}.

\subsection{Problem statement}
\label{subsec:problem_statement}

The input to the GPU recommendation tool consists of: an LLM model $M$ with unknown performance, a set of GPU profiles $\mathbb{G}$ that the user considers for deployment (each defined as the number and type of GPUs assigned to each pod), the latency constraints on nTTFT and ITL denoted as $L_1, L_2 \in \mathbb{R}^+$ respectively (and jointly denoted as $L=\left(L_1, L_2\right)$), and the expected load on the service expressed as the total number of concurrent users $U \in \mathbb Z^+$ that will be simultaneously sending requests to the service, following the same request distribution as modeled by the workload generator.
As we have argued in Sec. \ref{sec:intro}, an important assumption of the recommendation tool is to make no performance evaluations of the unseen LLM $M$.
Instead, LLM-Pilot uses the collection of historical performance data $\mathbb{D}_\text{train}$ collected using a set of training LLMs $\mathbb{M}_\text{train}$ on GPU profiles $\mathbb{G}$ to make predictions regarding nTTFT $l_1$ and ITL $l_2$ of the unseen LLM $M$.
In all equations to follow, we omit the dependence on the total number of users $U$ and the latency constraints $L$ for brevity.
The end goal of the GPU recommendation tool is to identify the most cost-effective GPU profile $G^\star \in \mathbb{G}$ and estimate the number $n$ of pods running on GPU profile $G^\star$ that should be created in order for LLM $M$ to serve $U$ concurrent users under constraints $L$:
\begin{equation}
	\text{find}\ \ G^\star\!\left(M\big|\mathbb{D}_\text{train}\right) = \text{arg} \min_{G \in \mathbb{G}} n\!\left(M, G\big|\mathbb{D}_\text{train}\right) \cdot c\!\left(G\right),
\label{eq:recommendation_problem}
\end{equation}

where
\begin{equation}
n\!\left(M, G\big|\mathbb{D}_\text{train}\right) = \ceil[\bigg]{\frac{U}{u_\text{max}\!\left(M, G\big|\mathbb{D}_\text{train}\right)}},\ \text{and}
\label{eq:calculate_n}
\end{equation}
\begin{equation}
    \begin{array}{l}
    u_\text{max}\!\left(M,\!G\big|\mathbb{D}_\text{train}\right) = \\
    = \max\Big\{\!u \in \mathbb{U}\!:\!\!
       \begin{array}{l}
        	l_1\!\left(M,\!G,\!u' \big|\mathbb{D}_\text{train}\right)\!\leq\!L_1 \\ 
        	l_2\!\left(M,\!G,\!u' \big|\mathbb{D}_\text{train}\right)\!\leq\!L_2
	\end{array}
	\!\forall_{u' \in \mathbb{U}:\ u' \leq u}\!\Big\}.
    \end{array}
\label{eq:calculate_u_max}
\end{equation}
Value $c\!\left(G\right) \in \mathbb{R}^+$ denotes the cost of a single pod running on the GPU profile $G$, read from the GPU pricing tables.
Value $n\!\left(M, G\big|\mathbb{D}_\text{train}\right)\!\!\in\!\!\mathbb{Z}^+$ denotes the number of~pods~with GPU profile $G$ needed to serve $U$ users under constraint $L$, predicted based on $\mathbb{D}_\text{train}$.
$\mathbb{U}\!\subset\!\mathbb{Z}^+$ is~the set of all considered numbers of concurrent users, while $u_\text{max}\!\left(M, G\big|\mathbb{D}_\text{train}\right) \in \mathbb{U}$ denotes the maximum number of users that can be served by a single pod of $M$ deployed on $G$ without violating constraints $L$, predicted based on $\mathbb{D}_\text{train}$.
Finally, $l_1\!\left(M,\!G,\!u \big|\mathbb{D}_\text{train}\right)\!,\ l_2\!\left(M,\!G,\!u \big|\mathbb{D}_\text{train}\right) \in \mathbb{R}^+$ respectively denote estimated nTTFT and ITL of $M$ deployed on $G$ and serving $u$ concurrent users, predicted using a regressor trained on $\mathbb{D}_\text{train}$.

\subsection{LLM-Pilot's solution}
\label{subsec:solution}

Before LLM-Pilot can decide on the most cost-effective GPU profile for an unseen LLM, it uses the performance model to make predictions regarding the unseen LLM's performance.
The performance model's training data $\mathbb D_\text{train}$ was created using the performance characterization dataset described in Sec. \ref{subsec:dataset}.
The performance model takes an input the features describing the LLM $M$, features describing the GPU profile $G$ and the number of concurrent users $u \in \mathbb{U}$.
As output, it predicts the latencies $l_1$ and $l_2$ of that inference service.
The recommendation tool uses these predictions to identify the most cost-effective GPU profile, following Eq. \eqref{eq:recommendation_problem}--\eqref{eq:calculate_u_max}.

\subsubsection{Feature engineering}
The LLMs are characterized by the following set of features: LLM type (e.g., t5, codegen2), whether the LLM has an encoder-decoder or decoder-only architecture, the number of parameters, layers, positions and heads, whether flash attention was used, the vocabulary size, parameters for relative attention (maximum distance and number of buckets) and the data type used for training.
The set of features characterizing the GPU profiles includes the number of GPUs, memory capacity and bandwidth, GPU architecture, number of Tensor/RT/CUDA cores, number of texture mapping units, raster operations pipelines, and streaming multiprocessors, TFLOPS for various data types, compute capability, interface generation, form factor (SXM vs. PCIe) and finally whether the GPU is connected using NVLink.
The LLM features listed above were explained in detail in \cite{Attention}, and the GPU features in \cite{Daniel}.

\subsubsection{Regressor}
Internally, the GPU recommendation tool of LLM-Pilot uses an XGBoost regressor \cite{xgboost}.
As the end goal of LLM-Pilot's recommendation tool is to identify cost-effective GPU profiles rather than to make accurate latency predictions, we introduce two modifications to the regression task which will ultimately improve the GPU recommendations.
Our first modification is to apply sample weights to our training data, in which the closer each data point's latency metrics are to the latency constraints, the higher weight is assigned to it.
Initially, we define weights based on nTTFT as follows:
\begin{equation}
	w_1(M, G, u|\mathbb D_M) = 1 - \frac{\big| \hat l_1\left(M, G, u|\mathbb D_M\right) - L_1\big|}{\max\limits_{v \in \mathbb{U}}\big| \hat l_1\left(M, G, v|\mathbb D_M\right) - L_1\big|}
\label{eq:weights_1}
\end{equation}
where $\hat l_1\left(M, G, u|\mathbb D_M\right)$ denotes the true nTTFT latency of $M$ deployed on $G$ with $u$ concurrent users, extracted from the known performance $\mathbb D_M$ of LLM $M$, as $M \in \mathbb D_\text{train}$.
The weights for the ITL latency $w_2(M, G, u|\mathbb D_M)$ are calculated in an analogous way, using the true ITL latency $\hat l_2\left(M, G, u|\mathbb D_M\right)$ and the constraint $L_2$.
With this formula for the sample weights, training data points have weights inversely proportional to how far their latency is from the respective latency constraint.
We combine both weights using arithmetic mean.
The intuition behind our use of the sample weights is as follows: as we are solving the GPU recommendation problem~\eqref{eq:recommendation_problem}, the regressor's purpose is to estimate the maximum number of concurrent users that can be served on a given GPU under the latency constraints.
Therefore, it has to make accurate latency predictions mainly for those numbers of users for which the latency metrics are near the constraints.

However, the sample weights can also lead to drastically incorrect recommendations.
For example, let us assume that for some GPU profile the true latency for 4 concurrent users is far from the latency constraint and therefore, that data point has a low sample weight.
The regressor predicts an incorrect, very high latency value for that data point and LLM-Pilot determines that the latency constraint is already violated. 
At the same time, the regressor made accurate latency predictions for 16 and 32 concurrent users, as these data points are close to the latency constraint and have high sample weights.
However, constraint violation for 4 concurrent users causes LLM-Pilot to determine that this GPU profile can only support 2 concurrent users and massively overestimate the number of pods needed.
Therefore, we additionally enforce on the regressor a monotonicity constraint on the number of concurrent users, as based on our experiments, as the number of concurrent users increases, the nTTFT and ITL of the service increase or stay constant:
\begin{equation*}
\begin{array}{c}
     u'\!<\!u'' \implies 
	l\!\left(M, G, u'\big|\mathbb{D}_\text{train}\right) \leq l\!\left(M, G, u''\big|\mathbb{D}_\text{train}\right) \\
	\hfill\forall_{u', u'' \in \mathbb{U},\ l \in \{ l_1,  l_2\}}
\end{array}
\end{equation*}

With the addition of the monotonicity constraint, if the regressor makes accurate predictions for data points close to the latency constraint, it will never incorrectly indicate violation or satisfaction of the constraint for the other data points.
This way, the monotonicity constraint ensures that using the sample weights does not negatively impact the estimation of the maximum number of concurrent users.

\begin{figure*}[t]
\centering
\begin{subfigure}[t]{.32\textwidth}
  \centering
  \includegraphics[width=\linewidth]{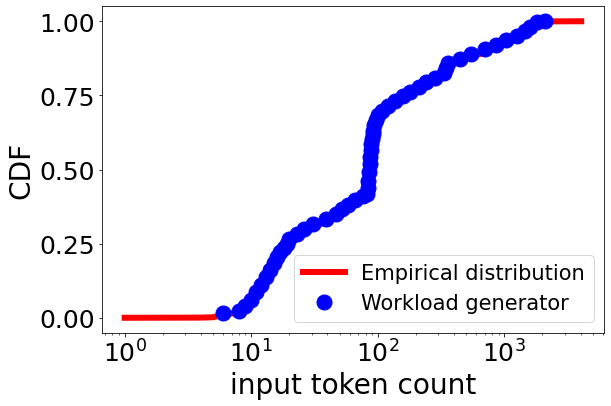}
  \caption{Number of input tokens}
  \label{subfig:histogram_a}
\end{subfigure} \hfill
\begin{subfigure}[t]{.32\textwidth}
  \centering
  \includegraphics[width=\linewidth]{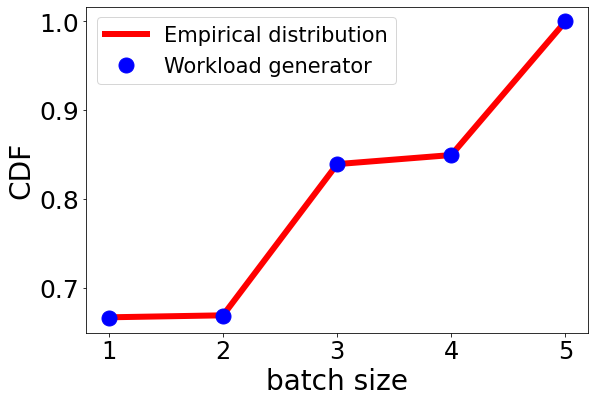}
  \caption{Client-side batch size}
  \label{subfig:histogram_b}
\end{subfigure} \hfill
\begin{subfigure}[t]{.32\textwidth}
  \centering
  \includegraphics[width=\linewidth]{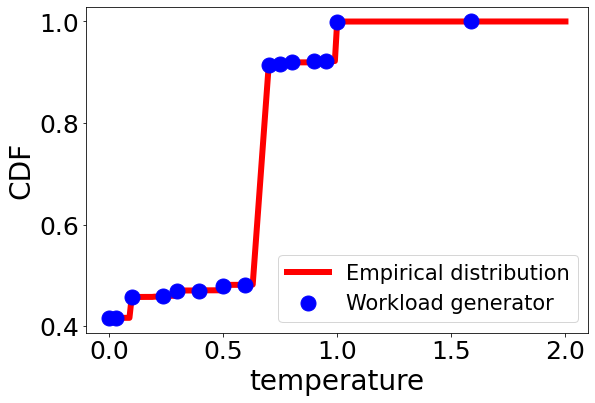}
  \caption{Temperature}
  \label{subfig:histogram_c}
\end{subfigure}
\caption{Marginal CDFs of selected request parameters in the empirical distribution and in the workload generator.}
\label{fig:histogram}
\end{figure*}

\begin{table*}[t]
\centering
\caption{LLMs and GPUs included in our performance characterization dataset: combinations for which data was collected (\greencheck), combinations in which the GPU profile's memory capacity was too small to host the LLM while leaving sufficient space to process workload generator's requests (\redcross), and combinations omitted due to software or hardware limitations (\yellowdash).}
\begin{tabular}{|l|c|c|c|c|c|c|c|c|c|c|c|c|c|c|c|}
\hline
\multicolumn{2}{|c|}{\multirow{2}{*}{LLM}} & \multicolumn{3}{c|}{H100 (80GB)}     & \multicolumn{3}{c|}{A100 (40GB)} & \multicolumn{2}{c|}{\!A10 (24GB)\!} & \multicolumn{3}{c|}{T4 (16GB)} & \multicolumn{3}{c|}{V100 (16GB)} \\\cline{3-16}
\multicolumn{2}{|c|}{} & {\ \ \!1\!\ \ } & {\ \ \!2\!\ \ } & {\ \ \!4\!\ \ } & {\ \ \!1\!\ \ } & {\ \ \!2\!\ \ } & {\ \ \!4\!\ \ } & {\ \ \!1\!\ \ } & {\ \ \!2\!\ \ } & {\ \ \!1\!\ \ } & {\ \ \!2\!\ \ } & {\ \ \!4\!\ \ } & {\ \ \!1\!\ \ } & {\ \ \!2\!\ \ } & {\ \ \!4\!\ \ } \\\hline
google/flan-t5-xl \cite{xl}              &  3B         & \greencheck & \greencheck   & \greencheck  & \greencheck & \greencheck  & \greencheck  & \greencheck  & \greencheck  & \greencheck  & \greencheck  & \greencheck  & \greencheck  & \greencheck  & \greencheck{}  \\\hline
google/flan-t5-xxl \cite{xl}             &  11B       & \greencheck & \greencheck   & \greencheck  & \greencheck & \greencheck  & \greencheck  & \redcross      & \greencheck  & \redcross      & \redcross      & \greencheck  & \redcross      & \redcross      & \greencheck{}  \\\hline
google/flan-ul2 \cite{ul2}                 &  20B       & \greencheck & \greencheck   & \greencheck  & \redcross     & \greencheck  & \greencheck  & \redcross      & \redcross      & \redcross      & \redcross      & \redcross      & \redcross      & \redcross      & \redcross{}      \\\hline
ibm/mpt-7b-instruct2 \cite{mpt}        &  7B         & \greencheck & \yellowdash       & \yellowdash      & \greencheck & \yellowdash      & \yellowdash      & \redcross      & \yellowdash      & \redcross      & \yellowdash      & \yellowdash      & \redcross      & \yellowdash      & \yellowdash{}      \\\hline
bigscience/mt0-xxl \cite{mt0}            &  13B       & \greencheck & \yellowdash       & \yellowdash      & \greencheck & \yellowdash      & \yellowdash      & \redcross      & \yellowdash      & \redcross      & \yellowdash      & \yellowdash      & \redcross      & \yellowdash      & \yellowdash{}      \\\hline
Salesforce/codegen2-16B \cite{codegen}   &  16B       & \greencheck & \yellowdash       & \yellowdash      & \redcross     & \yellowdash      & \yellowdash      & \redcross      & \yellowdash      & \redcross      & \yellowdash      & \yellowdash      & \redcross      & \yellowdash      & \yellowdash{}      \\\hline
Llama-2-7b \cite{Llama}                          &  7B         & \greencheck & \greencheck   & \greencheck  & \greencheck & \greencheck  & \greencheck  & \greencheck  & \greencheck  & \redcross      & \greencheck  & \greencheck  & \yellowdash      & \yellowdash      & \yellowdash{}      \\\hline
Llama-2-13b \cite{Llama}                        &  13B       & \greencheck & \greencheck   & \greencheck  & \greencheck & \greencheck  & \greencheck  & \redcross      & \greencheck  & \redcross      & \redcross      & \greencheck  & \yellowdash      & \yellowdash      & \yellowdash{}      \\\hline
EleutherAI/gpt-neox-20b \cite{neox}   &  20B       & \greencheck & \greencheck   & \greencheck  & \redcross     & \greencheck  & \greencheck  & \redcross      & \greencheck  & \redcross      & \redcross      & \greencheck  & \yellowdash      & \yellowdash      & \yellowdash{}      \\\hline
bigcode/starcoder \cite{starcoder}             &  15B       & \greencheck & \greencheck   & \greencheck  & \greencheck & \greencheck  & \greencheck  & \redcross      & \greencheck  & \redcross      & \redcross      & \greencheck  & \yellowdash      & \yellowdash      & \yellowdash{}      \\\hline
\end{tabular}
\label{tab:dataset}
\end{table*}

\subsubsection{Hyperparameter tuning}
\label{subsubsec:hpt}
Before training an XGBoost regressor, one must first set a number of \emph{hyperparameters} (HPs), which cannot be tuned as part of the training process but strongly impact the quality of the regressor's predictions.
For XGBoost, these include the number of boosted trees, their maximum depth, learning rate, subsampling rates, the tree building method, and the number of bins for the histogram tree method.
We tune XGBoost's HPs via a leave-one-LLM-out cross-validation procedure.
We split the available performance data into the training dataset $\mathbb D_\text{train}$ used to train the regressor, and the validation dataset $\mathbb D_\text{val}$ used to evaluate the predictions.
Specifically, all performance data from one LLM is used as $\mathbb D_\text{val}$ and all remaining LLMs act as $\mathbb D_\text{train}$.
Finally, we select the configuration of HPs that achieved the lowest average validation error across all possible training/validation splits.
As the error metric, we use the mean absolute percentage error (MAPE) weighted using the sample weights defined in Eq. \eqref{eq:weights_1} because it measures the error relative to the latency values, which vary significantly within our data.

\subsubsection{Final GPU recommendation}
Once the performance model has predicted the latencies for an unseen LLM $M$ across all GPU profiles and numbers of concurrent users, LLM-Pilot recommends the most cost-effective GPU profile and the number of pods that will safisfy the performance requirements, following Eq. \eqref{eq:recommendation_problem}--\eqref{eq:calculate_u_max}.
Then, LLM-Pilot can be used to tune the maximum batch weight for that LLM on that GPU profile and to deploy the inference service, as described in Sec. \ref{subsec:load_testing}.

%% file: 5-evaluation.tex
% !TEX root = main.tex
\section{Analysis and Evaluation}
\label{sec:evaluation}

In this Section we analyze and evaluate LLM-Pilot's workload generator (Sec. \ref{subsec:eval_generator}), the performance dataset collected using the performance characterization tool (Sec. \ref{subsec:dataset}), and the GPU recommendation tool (Sec. \ref{subsec:evaluation}).

\subsection{Workload  generator}
\label{subsec:eval_generator}

To evaluate the workload generator developed in this work (Sec. \ref{subsec:workload_generator}), we analyze: (1) whether the generator's internal model of requests accurately models the distributions of all parameters describing the traces, (2) whether the preserved correlation between request parameters has any effect on the inference performance, and (3) whether using the workload generator has any benefits over generating requests by drawing random samples from the traces.

\subsubsection*{Accurate modelling}
In Fig. \ref{subfig:histogram_a}-\ref{subfig:histogram_c} we compare the empirical marginal cumulative distribution function (CDF) of selected request parameters to the marginal CDF obtained with the workload generator. 
Based on the plots we can conclude that the workload generator preserves the marginal distributions of parameters with both very high and low cardinality.

\subsubsection*{Parameter correlation}
We have conducted an experiment for an example test case (Llama-2-13b running on one A100 80GB GPU) to prove that the correlation between request parameter affects the LLM inference performance.
On average, across 1--128 concurrent users, generating parameter values from independent marginal distributions results in 13\% lower throughput (up to 19\%), 30\% higher median TTFT (up to 98\%) and 25\% lower median ITL (up to 58\%), as compared to generating them based on a joint distribution.
We consider these differences significant enough to justify that modelling request parameters jointly is crucial for the measured performance to reflect what would be observed in production.

\subsubsection*{Size and sampling speed}
Because the parameters are strongly correlated, many combinations of parameters never occur, and their respective multi-dimensional bins are empty.
Consequently, the collection of multi-dimensional bins in the workload generator is sparse, with 46.5 thousand non-empty bins, compared to 10.7 billion theoretically possible combinations of parameter bin assignments.
Thanks to binning the parameter values and sparsity, the workload generator is much smaller in terms of storage space than the traces that it models ($<$1MB generator, compared to 1.6GB of traces), and will remain approximately the same size even if a much larger amount of traces is collected in the future.
Furthermore, the workload generator produces requests much faster than directly sampling past requests from the traces.
For an example of drawing 1000 samples, sampling requests from traces takes 770ms, while the workload generator produces the requests in 22ms, which is 35$\times$ faster.
It is worth noting that the time needed to generate 1000 requests is lower than the typical ITL of generating a single output token.

\subsection{Performance characterization}
\label{subsec:dataset}

Using LLM-Pilot's performance characterization tool, we performed a series of performance measurements using 10 LLMs deployed on 14 GPU profiles (single-GPU or tensor-parallel deployment across 2 or 4 GPUs, with one of 5 GPU types).
Table \ref{tab:dataset} presents which combinations of LLMs and GPU profiles were feasible.
For certain combinations it was impossible to collect performance data because the LLM would not fit into the GPU profile's aggregate memory or because the free space after loading the LLM into memory was insufficient to process the largest requests produced by the workload generator.
Additionally, certain combinations were impossible due to software or hardware constraints.
Specifically, at the time of writing this work TGIS didn't support tensor parallelism for certain LLMs.
Furthermore, TGIS uses flash attention \cite{flashattention} for some LLMs and therefore these LLMs couldn't be deployed on the V100 GPUs because of insufficient CUDA capability.

\begin{figure}[t]
\centering
\begin{subfigure}[t]{.45\textwidth}
  \centering
  \includegraphics[width=\linewidth]{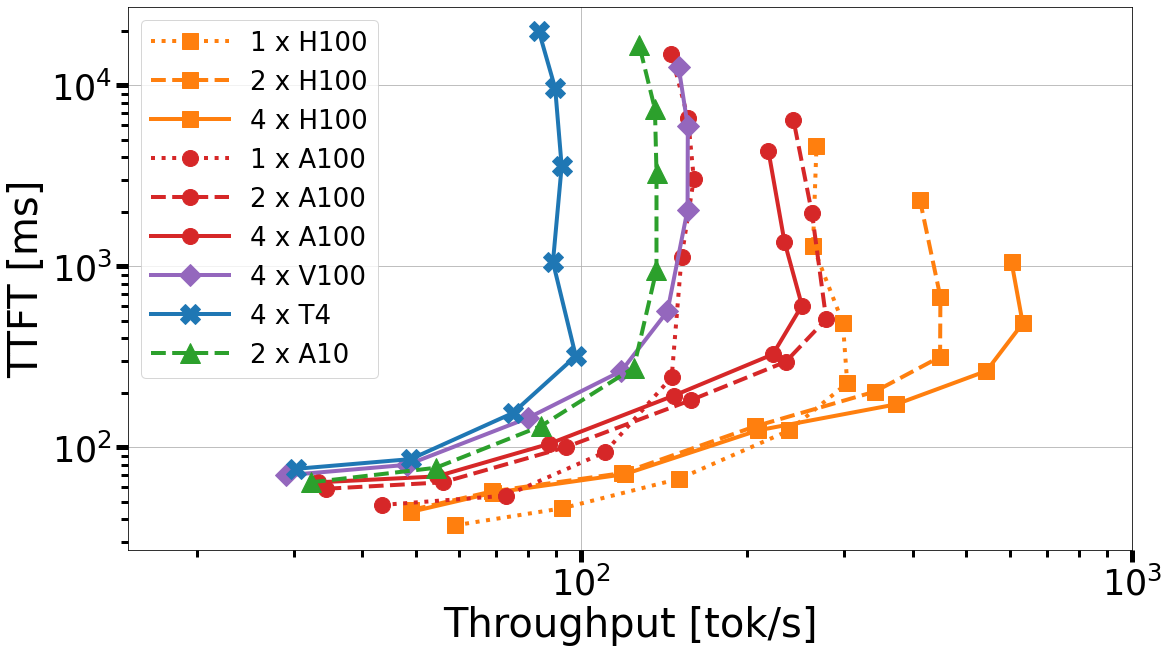}
  \caption{TTFT vs. throughput}
  \label{subfig:latency_v_throughput_a}
\end{subfigure}\hfill
\begin{subfigure}[t]{.45\textwidth}
  \centering
  \includegraphics[width=\linewidth]{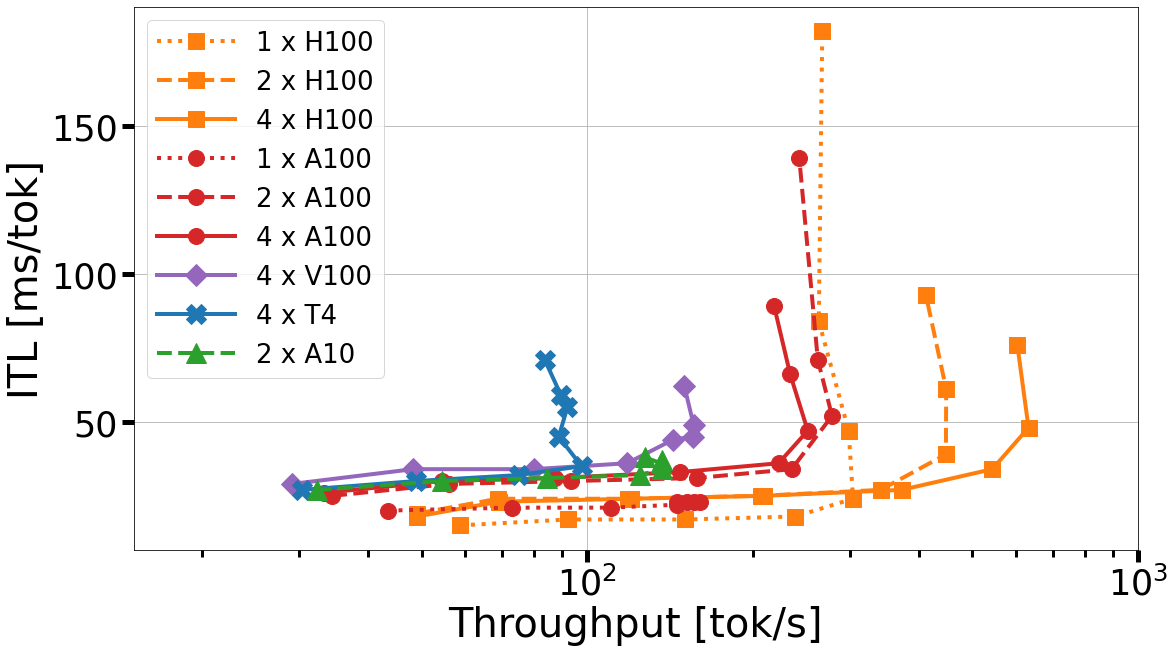}
  \caption{ITL vs. throughput}
  \label{subfig:latency_v_throughput_b}
\end{subfigure}\hfill
\begin{subfigure}[t]{.45\textwidth}
  \centering
  \includegraphics[width=.98\linewidth]{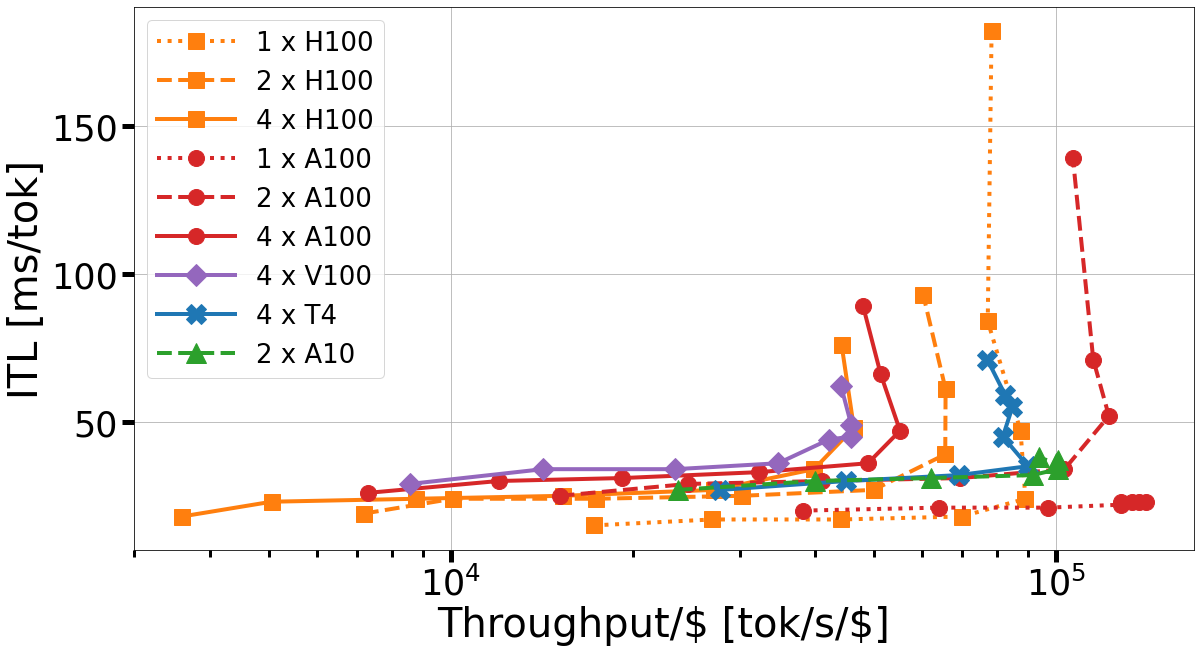}
  \caption{ITL vs. throughput per cost}
  \label{subfig:latency_v_throughput_c}
\end{subfigure}
\caption{Relationship between TTFT, ITL, throughput, and throughput per dollar for google/flan-t5-xxl LLM \cite{xl} running on a variety of GPUs. Markers on each curve mark exponentially increasing numbers of users $\in \{1, 2, 4, ..., 128\}$.}
\label{fig:latency_v_throughput}
\end{figure}

\subsubsection*{Performance data analysis}
In Fig. \ref{subfig:latency_v_throughput_a}-\ref{subfig:latency_v_throughput_b} we present the relationship between median latencies (TTFT and ITL) and throughput achieved by an example LLM deployed on various GPU profiles.
As the prompt processing phase is compute bound \cite{phase_bounds}, the typical behavior that we observe across all LLMs is that TTFT increases linearly with the increasing number of concurrent users because the LLM processes a larger batch of requests at the same time.
For weak GPUs with many concurrent users, we can observe a sudden jump of TTFT due to increased queueing time.
On the other hand, the decode phase is memory bandwidth bound \cite{phase_bounds}.
As a result, ITL typically remains stable as the number of concurrent users and throughput increase, until the memory capacity is saturated.
As the number of concurrent users increases further, the ITL increases rapidly, while throughput does not improve anymore.
We can also observe that the larger the total memory capacity of the GPU profile, the larger number of concurrent users marks the point of saturation.
Consequently, GPU profiles with larger memory capacity (and larger maximum batch weights) achieve higher throughput and lower ITL before saturation.

Additionally, in Fig. \ref{subfig:latency_v_throughput_c} we present the relationship between median ITL and throughput per dollar for the same LLM.
Throughput per dollar is defined as throughput divided by the cost of the respective GPU profile, and can help quantify the trade-off between the inference performance achieved using various GPU profiles and the cost that they incurred.
As the cost metric, we use hourly on-demand GPU instance prices collected from Amazon Web Services (AWS) pricing tables.
However, the user of LLM-Pilot could also plug in their own pricing table or use any other metric to express their preference for certain GPUs.
The plot demonstrates that the GPU profiles with the highest memory capacity are not necessarily the most cost-effective.
For example, although profiles with H100 GPUs outperform others in terms of maximum throughput, profiles using A100 and T4 GPUs achieve higher throughput per dollar.
However, if the service is subject to a very low latency constraint, many GPU profiles will not deliver on the SLA, even with just one concurrent user.
In such cases, it is necessary to use GPU profiles with higher memory capacity despite their high cost.

To the best of our knowledge, the dataset that we have collected is the first work comparing the performance of many LLMs on a variety of GPUs (discussed in Sec. \ref{subsec:benchmarking_tools}).
It is also the first dataset in which the maximum batch weight was optimized individually for each GPU profile.
We have made our performance characterization dataset public in the hope that it will benefit the community and extend the efforts towards maximizing the performance of LLM inference.

\subsubsection*{Characterization overhead}
We expect the characterization tool to be used whenever users wish to add support for new LLMs or GPUs, which could happen frequently if support for many novel LLMs is desired.
We estimate that collecting a performance dataset of similar size to ours would take approx. 8h: 5h to tune the maximum batch weights for all LLMs (30min/LLM, parallelized over GPUs), and 3h to run load testing experiments (20min/LLM, parallelized over GPUs).
We note that the characterization tool was designed to be used offline by the cluster administrator, so although its overhead is nonnegligible, it does not disrupt the use of the GPU recommendation tool.

\subsection{GPU recommendation}
\label{subsec:evaluation}

We evaluate LLM-Pilot's GPU recommendation tool, and multiple state-of-the-art performance prediction methods, using the performance characterization dataset described in Sec.~\ref{subsec:dataset}.
We simulate a set of ``unseen'' LLMs, $\mathbb{M}_\text{test}$, via a nested cross-validation procedure, i.e., by iteratively excluding one LLM $M$ from the performance characterization dataset and assuming its performance is unknown. We then tune the HPs of each method using the cross-validation procedure described in Sec. \ref{subsubsec:hpt} using only the performance data collected from the remaining models. 
We then use the regressors with tuned HPs to make performance predictions for $M$, and make recommendations.
In the following experiments, we assume that the required total number of concurrent users $U = 200$, the latency constraints on nTTFT and ITL are equal $L_1 = 100ms$ and $L_2 = 50ms$, respectively, and the possible numbers of concurrent users per pod $u \in \mathbb{U}=\{1, 2, 4, 8,... 128\}$.

\subsubsection*{Evaluation metrics}
In order to evaluate LLM-Pilot's ability to solve problem \eqref{eq:recommendation_problem}, we have defined three evaluation metrics.
The first evaluation metric is \textbf{success rate} $\mathcal S$.
The recommendation for LLM $M \in \mathbb{M}_\text{test}$ is considered a success $\mathcal S_M$ if, after the user has followed the tool's suggestion and deployed $n$ pods with the GPU profile $G^\star\!\left(M\big|\mathbb{D}_\text{train}\right)$ (denoted as $G^\star_M$ for brevity), the true performance of the inference service did in fact meet their initial requirements regarding the total number of concurrent users $U$ and latency constraints $L$:
\begin{equation}
\mathcal S_M\!=\!\Bigg\{\!\!\!
       \begin{array}{l}
        	1 \ \text{if}\ n\!\left(M, G^\star_M\big|\mathbb{D}_\text{train}\right)\!\cdot\!\hat u_\text{max}\!\left(M, G^\star_M\big|\mathbb{D}_M\right) \geq U, \\ 
        	0 \ \text{otherwise.}
	\end{array}
\label{eq:success}
\end{equation}
where $\hat u_\text{max}\!\left(M,\!G^\star_M\big|\mathbb{D}_M\right)$ denotes the true maximum number of concurrent users that can be served within a single pod with the GPU profile $ G^\star_M$ without violating the latency constraints $L$ which could be determined if the real performance data $\mathbb{D}_M$ of LLM $M$ was known.
To balance the success rate, in successful cases $\mathbb{M}_\text{test}^\mathcal{S} = \{ M \in \mathbb{M}_\text{test}: \mathcal S_M=1 \}$ we additionally calculate the \textbf{relative overspend} $\mathcal{O}_M \in \mathbb{R}^+$, which quantifies the relative difference between the expense that the user carried by following the tool's recommendation ($n$ pods with GPU profile $G^\star_M$ determined by LLM-Pilot) and the expense of the truly most cost-effective deployment (denoted as $\hat n$ pods with the GPU profile $\hat G^\star_M$) that they could have chosen if they knew the real performance $\mathbb{D}_M$ of LLM $M$:
\begin{equation}
\mathcal{O}_M = \frac{n\big(M,\! G^\star_M\big)\!\cdot\!c\!\left(G^\star_M\right) - \hat n\big(M,\!\hat G^\star_M\big|\mathbb{D}_M\big)\!\cdot\!c\big(\hat G^\star_M\big)}{\hat n\big(M,\!\hat G^\star_M\big|\mathbb{D}_M\big)\!\cdot\!c\big(\hat G^\star_M\big)}.
\end{equation}
We obtain the final success rate $\mathcal{S} \in \left[ 0, 1 \right]$ by averaging the successes of all unseen LLMs $\mathbb{M}_\text{test}$, and the mean relative overspend $\mathcal{O} \in \mathbb{R}^+$ by averaging the relative overspends over all unseen LLMs for which the recommendation was successful $\mathbb{M}_\text{test}^\mathcal{S}$.
The purpose of using both metrics above is for them to compliment each other: the success rate penalizes underpredicting the number of GPUs needed, while the overspend penalizes overpredicting them.
Finally, we define the \textbf{S/O score} $\mathcal{SO} \in \left[ 0, 1 \right]$, which combines the success rate $\mathcal{S}$ and the inverse of overspend $\mathcal{O}$ using the harmonic mean:
\begin{equation}
\mathcal{SO} = \frac{2 \cdot \mathcal{S} \cdot \text{max}\!\left(0, 1 - \mathcal{O} \right)}{\mathcal{S} + \text{max}\!\left(0, 1 - \mathcal{O}\right)}.
\label{eq:so_score}
\end{equation}
The S/O score serves as the most important metric in our study, as it directly evaluates how well we solve problem \eqref{eq:recommendation_problem}.

\subsubsection*{Baselines}
While to the best of our knowledge there are no prior works that predict the inference performance of LLMs across various GPUs, we have implemented several methods developed in related fields. 
\textbf{PARIS} \cite{Yadwadkar} predicts the performance of a previously unseen application across many virtual machine (VM) types in the cloud.
First, PARIS measures the performance of the unseen application on two reference VM types: the weakest and the most powerful one.
Then, it uses a RF regressor to predict the inference performance of that application running on other VM types, based on the features describing the application and the performance measurements collected on the reference VM types.
In our implementation of PARIS, the performance measurements consist of nTTFT, ITL and throughput values for all numbers of concurrent users for two reference GPU profiles: 1$\times$T4 and 4$\times$H100, which, respectively, have the weakest and the strongest memory and computing parameters.
To evaluate how the performance measurements of the reference VM types improve the quality of RF predictions in PARIS, we have also implemented a \textbf{RF}~predictor that uses as input only features describing the LLM, and makes no performance measurements.
\textbf{Selecta} \cite{Selecta} was developed for the same use-case as PARIS.
Internally, it builds a sparse matrix containing the known runtimes of historical and reference (application, VM type)-combinations, and predicts the missing entries via collaborative filtering. 
We have implemented Selecta using the same library as the original work \cite{surprise}, and have chosen the same reference GPU profiles as we used for PARIS.
\textbf{Morphling} \cite{Morphling}, \textbf{PerfNet} \cite{PerfNet}, and \textbf{PerfNetV2} \cite{PerfNetV2} predict the performance of inference services using neural network (NN) models. 
We have implemented all three NNs ourselves.
As Morphling additionally fine-tunes the NN using two reference evaluations, we again use the same reference GPU profiles as for other baselines.
Finally, \textbf{Static policy} is a simple, naive baseline, in which no performance predictions are made.
The GPU recommendation is always the same: to create a fixed number of pods with a certain GPU profile.
We have considered a broad range of static policies and present the one which achieved the highest S/O score: 4 pods running on 1$\times$A100 GPU.
Unknown HPs of baseline methods were determined through the same leave-one-LLM-out cross-validation procedure as for LLM-Pilot.

\begin{figure}[t]
\centerline{\includegraphics[width=\columnwidth]{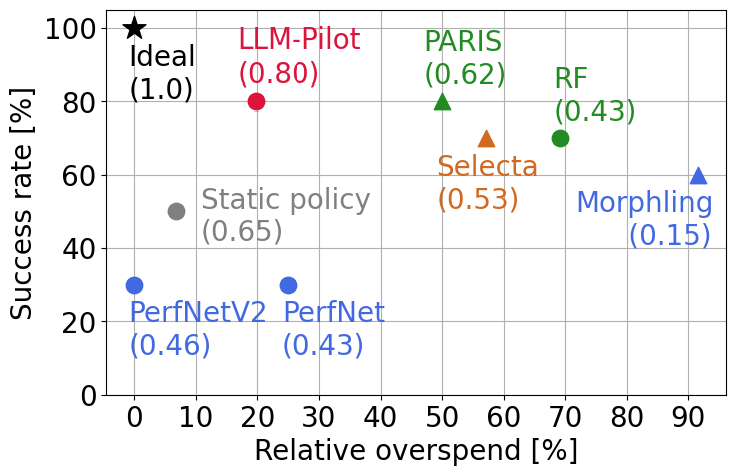}}
\caption{Evaluation of the quality of recommendations made by LLM-Pilot and baselines. Numbers in parentheses present the S/O scores achieved by each method. We mark methods which make reference performance measurements using $\blacktriangle$, and methods which make no reference evaluations with $\CIRCLE$. Additionally, $\bigstar$ marks the theoretical ideal performance.}
\label{fig:results}
\end{figure}

\subsubsection*{Recommendation results}
Fig. \ref{fig:results} summarizes the recommendation scores achieved with LLM-Pilot and the baseline methods.
Based on these results, we can draw a number of conclusions.
Firstly, LLM-Pilot achieves good results -- its recommendations are successful in 80\% of cases and have average overspend of less than 20\%, outperforming all other methods in terms of the S/O score.
The static policy can be considered a high-risk, high-reward solution.
It  only succeeds in 50\% of cases but it achieves an excellent overspend when it makes a successful recommendation, outperforming all other baselines in terms of the S/O score.
PARIS and Selecta achieve the same or similar success rate as LLM-Pilot but have higher average overspend, and require making additional performance measurements using the reference GPU profiles. 
RF, which depicts the performance of PARIS without the reference performance measurements, is significantly worse in terms of all three evaluation metrics. 
Finally, PerfNet models achieve good overspend scores but their success rate is the worst of all state-of-the-art solutions.
On the other hand, Morphling achieves higher success rate than other NN-based methods thanks to performing reference performance measurements but is worse in terms of overspend.

Overall, the experimental results indicate that LLM-Pilot can successfully recommend the most cost-effective GPU profile for previously unseen LLMs, and outperforms all state-of-the-art methods.
Its GPU recommendations satisfy the performance requirements 33\% more frequently on average, thanks to ensuring that the performance is most accurately predicted in the neighborhood of the latency constraints.
At the same time, it recommends GPU profiles which are on average 60\% cheaper than state-of-the-art.

%% file: 6-related_works.tex
% !TEX root = main.tex
\section{Related Works}
\label{sec:related_works}

\subsection{LLM traces}

There are multiple publicly available collections of LLM traces, in some cases consisting of thousands or millions of LLM inference requests.
While some datasets consist of real user inputs of inference requests and resulting outputs \cite{FlanCollection, ShareGPT, ChatbotArena}, they do not take into account any additional request parameters.
There is also a range of inference request collections generated synthetically: \cite{OpenAssistant} generated by a group of volunteers, \cite{OpenOrca, Orca2} with handcrafted system prompts, and \cite{Alpaca} generated by GPT 3.5 using Self-Instruct \cite{selfinstruct}.
None of the synthetic collections represents a real-life distribution of user requests, and therefore could not be used in this work for realistic workload generation.

\subsection{Workload generators and benchmarking tools}
\label{subsec:benchmarking_tools}

There is a number of related works on benchmarking LLM inference services, which we have compared in Table~\ref{tab:literature}.
Many of them measure the LLM inference performance under a very simple workload of requests not based on real LLM usage \cite{LLMPerf, Optimum, inference-benchmark}.
As we have argued in Sec. \ref{subsec:workload_generator}, a realistic and varied set of inference requests is necessary to perform meaningful performance measurements. 
Other benchmarking tools ensure that the inference requests are realistic by drawing random samples from existing trace collections \cite{vLLM, MLPerf, Fleece}.
However, none of the related benchmarking tools optimizes the maximum batch weight, which we have found to have big influence on the performance.

\subsection{LLM performance datasets}

While various benchmarking tools publish some of their benchmarking results, none of the existing datasets aggregates performance measurements of many LLM services deployed on a variety of GPUs (see Table \ref{tab:literature}).
Optimum's \cite{Optimum} LLM-Perf leaderboard \cite{LLMPerf-leaderboard} includes benchmarking results of LLMs but only includes 2 GPUs, while MLPerf \cite{MLPerf} collected data across many GPUs but only two LLMs are currently included.
Other related benchmarking tools \cite{Fleece, inference-benchmark, vLLM, LLMPerf} also published small collections of LLM performance data.

\begin{table}[t]
\centering
\caption{Comparison of LLM-Pilot's performance characterization tool and related LLM benchmarking tools, including LLM performance datasets that they released publicly.}
\begin{tabular}{|l|c|c|c|c|}
\hline
\multirow{2}{*}{\begin{tabular}[c]{@{}l@{}}Comparison\\criterion\end{tabular}} & \multirow{2}{*}{\begin{tabular}[c]{@{}c@{}}Workload\\based on\\real data\end{tabular}} & \multirow{2}{*}{\begin{tabular}[c]{@{}c@{}}Maximum\\batch weight\\tuning\end{tabular}} & \multicolumn{2}{c|}{\begin{tabular}[c]{@{}c@{}}LLM performance\\data released publicly\end{tabular}} \\ \cline{4-5}
 &  &  & \begin{tabular}[c]{@{}c@{}}Number\\of LLMs\end{tabular} & \begin{tabular}[c]{@{}c@{}}Number\\of GPUs\end{tabular} \\ \hline
Optimum~\cite{Optimum} & $\times$ & $\times$ & 34 & 2 \\ \hline
LLMPerf \cite{LLMPerf} & $\times$ & $\times$ & 3 & 1 \\ \hline
\begin{tabular}[c]{@{}l@{}}Inference\\benchmark~\cite{inference-benchmark}\end{tabular} & $\times$ & $\times$ & 1 & 1 \\ \hline
Fleece \cite{Fleece} & \checkmark & $\times$ & 5  & 5 \\ \hline
vLLM \cite{vLLM} & \checkmark & $\times$ & 3 & 2 \\ \hline
MLPerf \cite{MLPerf} & \checkmark & $\times$ & 2 & 10 \\ \hline
LLM-Pilot (ours) & \checkmark & \checkmark & 10 & 14 \\ \hline
\end{tabular}
\label{tab:literature}
\end{table}

\subsection{LLM performance prediction}

To the best of our knowledge, there are no prior works that predict the inference performance of LLMs across a variety of hardware platforms.
The works that are most closely related to our work predict the runtimes of other types of machine learning applications.
Many of these methods \cite{Yadwadkar, Morphling, Selecta, PerfNet, PerfNetV2} have been used as baselines in this work and have been discussed in detail in Sec. \ref{subsec:evaluation}.
Other works related to LLM-Pilot include \cite{Daniel} and \cite{Falcon} which predict the training runtime or inference latency of NNs across GPUs or runtime and resource configurations, while \cite{BIG-bench-paper} and \cite{Initializers} predict the quality of outputs of machine learning models on downstream tasks.

%% file: 7-conclusion.tex
% !TEX root = main.tex
\section{Conclusion and Next Steps}

In this work we have presented LLM-Pilot, a system that can perform realistic and optimized benchmarking of LLM inference services across different GPUs.
In addition, LLM-Pilot can recommend which GPU will meet performance requirements in the most cost-effective way for a previously unseen LLM, achieving on average 33\% higher success rate and 60\% lower cost compared to state-of-the-art methods.
As a next step, we intend to extend LLM-Pilot to cover the multi-tenancy scenario, in which multiple users compete to deploy LLM inference services on the same hardware resources.

%% file: acknowledgement.tex
% !TEX root = main.tex
\section*{Acknowledgement}
We would like to express our gratitude to our colleagues in IBM Research: Nick Hill, for sharing his technical insights into LLM serving using TGIS, and Burkhard Ringlein, for his valuable input during the experimental work and the process of writing this publication.